\documentclass[prd,twocolumn,showpacs,preprintnumbers,nofootinbib,amsmath,amssymb]{revtex4}

\usepackage{graphicx}
\usepackage{dcolumn}
\usepackage{bm}
\usepackage{amsmath}

\newcommand{\beq}{\begin{eqnarray}}
\newcommand{\eeq}{\end{eqnarray}}

\newcommand{\real}{{\sf I}\kern-.12em{\sf R}}
\newcommand{\comp}{{\sf I}\kern-.50em{\sf C}}
\newcommand{\unity}{{\sf I}\kern-.54em{\sf 1}}

\def\spose#1{\hbox to 0pt{#1\hss}}
\def\ltapprox{\mathrel{\spose{\lower 3pt\hbox{$\mathchar"218$}}
 \raise 2.0pt\hbox{$\mathchar"13C$}}}

\begin{document}

\title{Finite size phase transitions in QCD with adjoint fermions}
\author{Guido Cossu}
\affiliation{
Dipartimento di Fisica dell'Universit\`a di Pisa and INFN - Sezione di Pisa, I-56127 Pisa, Italy}
\email{g.cossu@sns.it}

\author{Massimo D'Elia}
\affiliation{Dipartimento di Fisica dell'Universit\`a di Genova and INFN - Sezione di Genova, I-16146 Genova, Italy}
\email{delia@ge.infn.it}

\date{\today}

\begin{abstract}
We perform a lattice investigation of QCD with three colors and
2 flavors of Dirac (staggered) fermions in the adjoint representation, 
defined on a $4 d$ space with one spatial dimension compactified,
and study the phase structure of the theory as a function 
of the size $L_c$ of the compactified dimension.
We show that four different phases take place, corresponding to 
different realizations of center symmetry: two center symmetric
phases, for large or small values of $L_c$, separated 
by two phases in which center symmetry is broken in two different
ways; the dependence of these results on the quark mass is discussed. 
We study also chiral properties and how they are affected by 
the different realizations of center symmetry; chiral symmetry, in particular,
stays spontaneously broken at the phase transitions and may be restored at much
lower values of the compactification radius.
Our results could be relevant to a recently proposed conjecture~\cite{KUY} of 
volume indepedence of QCD with adjoint fermions in the large $N_c$ limit.
\end{abstract}


\maketitle

\section{Introduction}
\label{introd}

Extracting the physics of large $N_c$ QCD from simulations or
analytic computations performed on small volumes, down to a single
site, has been 
the aim of large theoretical efforts since the original work
by Eguchi and Kawai~\cite{EK}. However, the expectation that physics
on a $d$-dimensional 
$L_1 \times \dots L_d$ periodic lattice should be independent,
as $N_c \to \infty$, of the lattice sizes $L_1, \dots L_d$, is 
justified if no transition occurs as the sizes $L_1, \dots L_d$ are changed:
this is not true in ordinary Yang-Mills theories in $d > 2$,
which undergo phase transitions as one or more dimensions are 
compactified below a given threshold, corresponding to 
the spontaneous breaking of the center symmetry in that 
direction. One or more
Wilson (Polyakov) lines in the compactified directions get a non-zero
expectation value, i.e. one or more deconfining-like transitions take
place (see Ref.~\cite{nn07} for a review on the subject).

Large $N_c$ QCD with $N_f$ fermions in the antisymmetric
representation, 
or QCD(AS), has attracted
lot of theoretical interest in the last few
years,
as a different $N_c \to \infty$ limit of ordinary QCD: it is 
indeed equivalent to ordinary QCD for $N_c = 3$.
Orientifold planar equivalence states that QCD(AS) is equivalent, in
the large $N_c$ limit and in the charge-even sector of the theory,
to QCD with fermions in the adjoint representation, QCD(Adj)~\cite{asv}.
Such equivalence is guaranteed if the charge conjugation symmetry 
is not spontaneously
broken in QCD(AS)~\cite{uy2006}: that is true for sufficiently large volumes,
but fails in presence of a compactified dimension, below a typical
compactification radius~\cite{deghof,lpp}.

Recently it has been made the hypothesis~\cite{KUY} 
that instead QCD with one or more adjoint fermions
would persist in the confined phase (i.e. no phase transition would occur)
as one of the dimensions is compactified, if
the boundary conditions for fermions are periodic
in that direction (i.e. if the compactified dimension
is not temperature-like). That would imply physics
be volume independent in the large $N_c$ limit, and 
thus would open the possibility of studying QCD(Adj)
down to arbitrarily small volumes, finally connecting
information, via orientifold planar equivalence, to the physical
properties of QCD(AS) on large volumes. That would
be a breakthrough in the study of large $N_c$ QCD,
with possible important implications also for
the knowledge of real QCD. It is therefore 
of great importance to work on such conjecture,
providing evidence from numerical 
lattice simulations. 

The argument given in Ref.~\cite{KUY} is based on the 
analysis of the 1-loop effective potential for the Polyakov
(Wilson) line, $\Omega$, taken along a compactified dimension of length $L_c$,
in presence of $N_f$ ($N_f/2$) flavors of massless Majorana (Dirac) 
adjoint fermions, which is given, up to a constant, by
\beq
   V_{\rm eff} (\Omega) = 
    \frac{1 \mp N_f}{24\pi^2L_c^4}
	\sum_{i, j=1}^{N_c} [\phi_i -\phi_j]^2  
( [\phi_i -\phi_j] - 2\pi)^2
\label{potential}
\eeq
where the $-/+$ sign occurs for periodic/antiperiodic boundary
conditions (p.b.c./a.b.c.) in the given direction, 
$e^{i \phi_j}$, $j = 1, \dots N_c$ are the eigenvalues of the
Polyakov line, which satisfy the constraint 
$\prod_j e^{i \phi_j} = 1$,
and $[x] = (x \bmod 2\pi)$.

At arbitrarily small compactified dimension, the 1-loop potential gives
all relevant information.
For $N_f = 0$ or for a.b.c., the
potential implies attraction among the eigenvalues: indeed it is 
minimized when they coincide, meaning that the Polyakov line alignes
itself along some center element and center symmetry is spontaneously
broken. Instead, for $N_f > 1$ and p.b.c.
the potential changes sign and becomes repulsive: the Polyakov line 
is disordered in this case and center symmetry is not broken, exactly
as in the infinite volume, confined phase. The potential in 
Eq.~(\ref{potential})
is given for zero quark mass $m$ and corrections are expected for
finite $m$, however results should not change as finite quark
masses are switched on, and indeed it has been shown in 
Ref.~\cite{myog09} that the disordered confined phase is always
preferred for small enough 
(compared to $\Lambda_{\rm QCD}$) quark masses.
Analogous results are obtained when studying the 1-loop
potential in the lattice regularized theory~\cite{Bringoltz},  
while contrasting results (i.e. a center symmetry breaking 1-loop potential) 
obtained in a three-dimensional reduced model~\cite{Bedaque}
should be interpreted with care and may be due to the absence of certain 
center stabilizing relevant operators which should in principle be included
in the reduced model~\cite{Bringoltz}.
Finally, a similar effect is obtained in the pure SU($N_c$) gauge theory
by adding a trace deformation (i.e. proportional to the 
adjoint trace of the Polyakov line) to the usual Yang-Mills 
action~\cite{UY08,myog07}.

An essential requirement for volume independence in the
large $N_c$ limit is that nothing,
i.e. no phase transition, happens as the compactified dimension spans
from infinity to the arbitrarily small, perturbative regime, i.e. that
the system stays always confined and does not explore any other phase
in between.
That can be verified by numerical lattice simulations. 
While numerical results exist for pure SU($N_c$) gauge theories with trace
deformations~\cite{myog07}, 
no result still exists concerning QCD with dynamical adjoint fermions.

In the present paper we consider the example of $N_c = 3$ with
2 flavors of Dirac adjoint fermions, providing numerical evidence
that the requirement above is highly non-trivial. 
The system that we have studied undergoes several
phase transitions when shrinking 
from the large volume to the small
volume confined regimes. Our study has been performed for finite
values of the quark masses: the fact that these phase transitions
may disappear in the chiral limit is not excluded, but our present
data show that this is non-trivial.

While our findings could be relevant just to the particular case
studied here 
and to the discretization setup used, they are 
a warning claiming for further studies with 
different values of $N_c$ and approaching the continuum limit.

The paper is organized as follows. In Section II we present 
details about lattice QCD with adjoint fermions and about
our numerical simulations. In Section III we present our 
numerical results, concerning the realization of both center symmetry
and chiral symmetry as a function of the size of the compactified dimension. 
In Section IV we discuss our conclusions.

\section{QCD(Adj) on the lattice and simulation details}

Fermions in the adjoint representation of $SU(3)$ have $8$ 
color degrees of freedom and can be described by $3\times 3$ hermitian
traceless matrices:
\begin{equation}
Q = Q^a\lambda_a
\end{equation}
where $\lambda_a$ are the Gell-Mann's matrixes. 
The elementary parallel trasports acting on them are given by the  
$8-$dimensional $U_{(8)}$ representation (which is real)
of the gauge links
\begin{equation}
U^{ab}_{(8)\ i,\mu} = \frac{1}{2}{\rm Tr}\left
(\lambda^aU_{(3)\ i,\mu}\lambda^bU_{(3)\ i,\mu}^{\dagger}\right )
\label{eq_u8}
\end{equation}
where $i$ is a lattice site. 
The full discretized lattice action used in our investigation is given by
\begin{equation}
S = S_G\left[U_{(3)}\right] + \sum_{i,j,a,b} {\bar Q}_i^a M\left[U_{(8)}\right
]_{i,j}^{a,b} Q_j^b
\end{equation}
$S_G$ is the standard pure gauge plaquette action with links in the
3-dimensional representation 
$$S_G  = \beta \sum_{\Box} \left(1 - \frac{1}{N_c}
{\rm Tr} \Pi_{\Box} \right) $$
where $\beta$ is the inverse gauge coupling
$\beta = {2 N_c}/{g_0^2}$ and the sum is over all 
elementary plaquette operators $\Pi_{\Box}$ of the lattice.
$M$ is the fermionic matrix, which has been chosen according to 
the standard staggered fermion discretization:
\begin{eqnarray}
M\left[U_{(8)}\right]_{i,j}^{a,b} &=& a m
\delta_{i,j} \delta^{a,b} 
+ \nonumber \\
{1 \over 2}  \sum_{\nu=1}^{4} && \hspace{-10pt} \eta_{i,\nu}\left(U_{(8)\ i,\nu}^{a,b} \delta_{i,j-\hat\nu}-
U^{\dag\ a,b}_{(8)\ i-\hat\nu,\nu} \delta_{i,j+\hat\nu}\right) 
\label{fmatrix}
\end{eqnarray}
Notice that, since gauge links acting on adjoint fermions are real
matrixes, $M_{i,j}^{a,b}$ is real as well.
The functional integral, describing 2 flavors of Dirac adjoint 
fermions in the continuum limit, is then given by
\beq
Z = \int \mathcal{D}U e^{-S_{G}} 
(\det M[U_{(8)}])^{1/2} 
\label{partfun1}
\eeq
Due to the reality of $M$, that can be expressed in terms of a real
pseudofermion field living only on even sites of the lattice as follows
\beq
Z = \int \mathcal{D}U \mathcal{D}\Phi_e 
e^{-S_{G}} \exp\left( - {\Phi^t_e} (M^t M)_{ee}^{-1} \Phi_e \right)  
\label{partfun2}
\eeq
In this form the functional integral distribution can be easily sampled
by a standard Hybrid Monte Carlo algorithm. In particular we have 
adopted the so-called $\Phi$ algorithm~\cite{Phi}.

We shall consider in the following a system in which one dimension
is compactified and has size $L_c$. Adjoint parallel transports
are blind to the center of the gauge group, hence the presence of
adjoint fermions does not break explicitely center symmetry in 
the compactified direction. The fundamental Polyakov line  
$L_{(3)}$ taken along the compactified dimension is not 
invariant under the same symmetry, and is therefore
an exact order parameter for its spontaneous breaking.

If fermions are given a.b.c.
in the compactified dimension, the Feynman path
integral can be interpreted as the partition function 
at temperature $T = 1/L_c$. The spontaneous breaking of center
symmetry is then associated to deconfinement. 
Various numerical studies of finite temperature QCD(Adj) have 
been performed in the past, mostly aimed at studying the relation between
deconfinement and chiral symmetry restoration~\cite{Karsch,Engels,Cossu}.

In our investigation we have considered fermions with p.b.c.
along the compactified dimension.
This theory cannot be interpreted as a thermodynamical system 
(apart from the case 
of infinite fermion mass), but must considered
simply as a system with a shorter, compactified spatial dimension. 
Nevertheless
we shall refer to the phase where the fundamental Polyakov line
acquires a non-zero expectation value, oriented along an 
element of the center of the gauge group, as a ``deconfined'' phase.

In most of our numerical simulations we have kept a fixed 
number of lattice sites along the compactified direction,
and varied its physical size by tuning the lattice spacing 
$a$ via the inverse gauge coupling $\beta$. Asymptotic 
freedom implies that $a \to 0$ as $\beta \to \infty$, hence
the compactified dimension is made shorter and shorter as 
$\beta$ is increased. Since we are mostly interested in studying
which phases are explored by the system as the size of the
compactified dimension is changed from large to small values, and not in 
the exact location of the possible phase transitions, it is enough 
for our purposes to look at the phase structure of theory
as a function of $\beta$ at fixed number of lattice sites.
However we shall try to get an estimate of physical scale ratios,
by means of the 2-loop perturbative $\beta$-function, describing
the asymptotic dependence of the lattice spacing $a$ on $\beta$,
or alternatively by performing numerical simulations also 
at a fixed 
value of the lattice spacing, in which $L_c$ is tuned by changing 
the number of 
lattice sites in the compactified dimension.

It is essential to repeat our investigation
for different values of the quark mass. An infinite quark
mass corresponds to the pure gauge theory, where it is well known that
two different phases, deconfined or confined, 
take place respectively for short or large values of the compactified
dimension; such phases are separated by a first order transition~\cite{Fukugita}.
For small enough quark masses one expects instead center symmetric
phases both for a very large or a very short size of the compactified dimension,
with or without different phases in between. It is then clear
that a highly non-trivial phase structure must be found in the
quark mass - $L_c$ plane. 
For this reason we have performed numerical
simulations at five different values of the bare quark mass,
$am = 0.50$, $0.10$, $0.05$, $0.02$ and $0.01$. 

In all cases
we have chosen molecular dynamics trajectories of length
$0.5$, with typical integration steps $\delta t$ ranging
from 0.025 (for $am = 0.50$) to 0.005  (for $am = 0.01$), so
as to keep an acceptance rate around $80\%$. For each parameter
choice ($am$, $\beta$, $L_c$) we have collected between 2k and 20k 
trajectories.
Numerical simulations have been performed on 
the APEmille machine in Pisa and the apeNEXT facility in 
Rome.

\section{Numerical results}

We start by presenting results obtained on a lattice $16^3 \times \hat L_c$,
with $\hat L_c = 4$ and p.b.c. for all fields. 
The shorter direction plays the role of the compactified dimension of 
length $L_c = \hat L_c a (\beta, m)$. 
In order to distinguish the various phases corresponding to 
different possible realizations of center symmetry, we have 
studied the expectation value and distribution 
of the trace of the Polyakov line, $L_{(3)}$, 
taken along the shorter direction and averaged over the orthogonal
$3 d$ space, as a function of $\beta$ and $am$.

\begin{figure}[h!]
\vspace{0.5cm}
\includegraphics*[width=1.0\columnwidth]{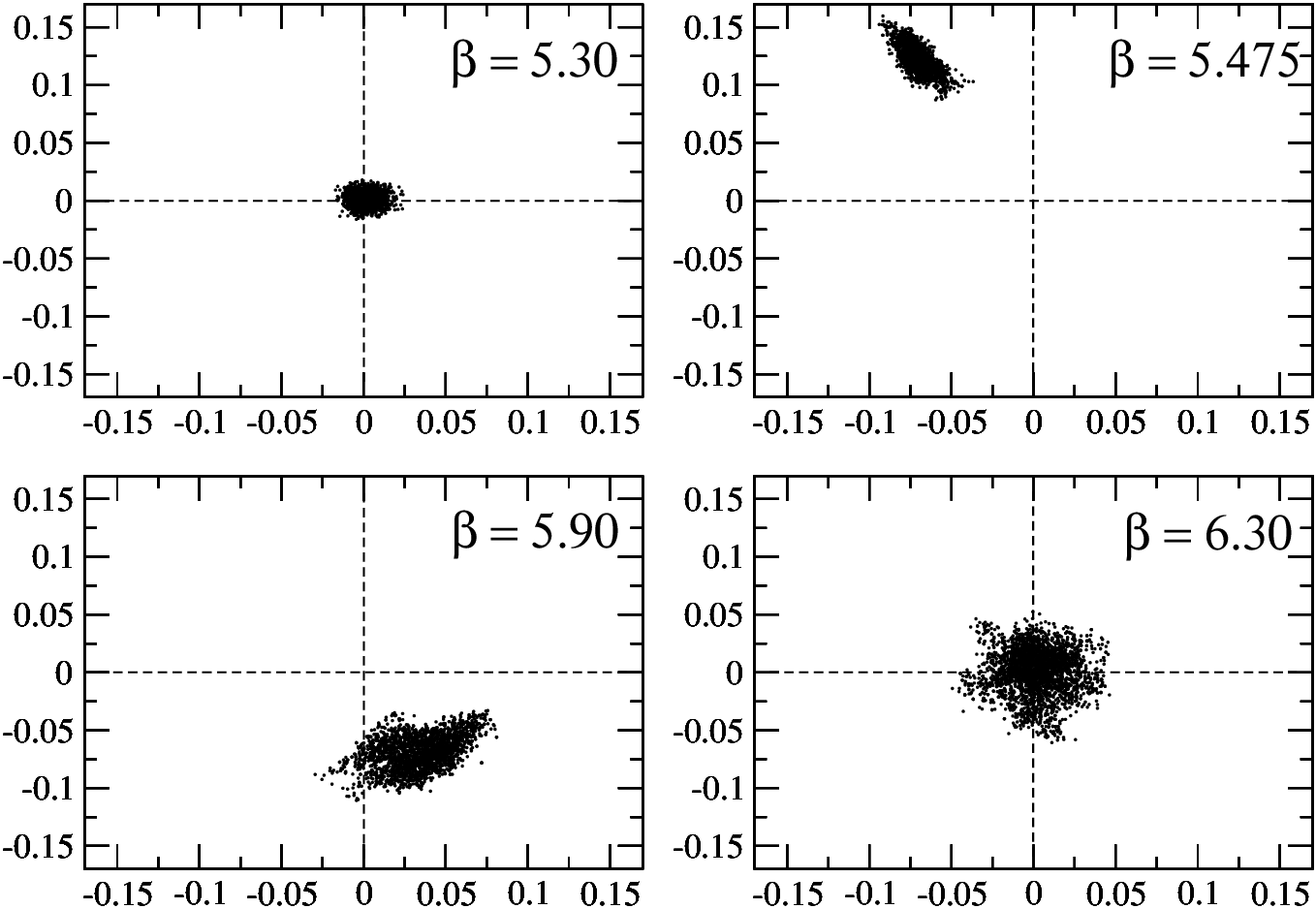}
\caption{Scatter plots of the distribution of the trace of the 
Polyakov line, $L_{(3)}$,
in the complex plane, for various 
values of $\beta$ and at a fixed quark mass $a m = 0.10$
on a $16^3 \times 4$ lattice.
}
\label{distr1} 
\end{figure}

\begin{figure}[h!]
\vspace{0.5cm}
\includegraphics*[width=1.0\columnwidth]{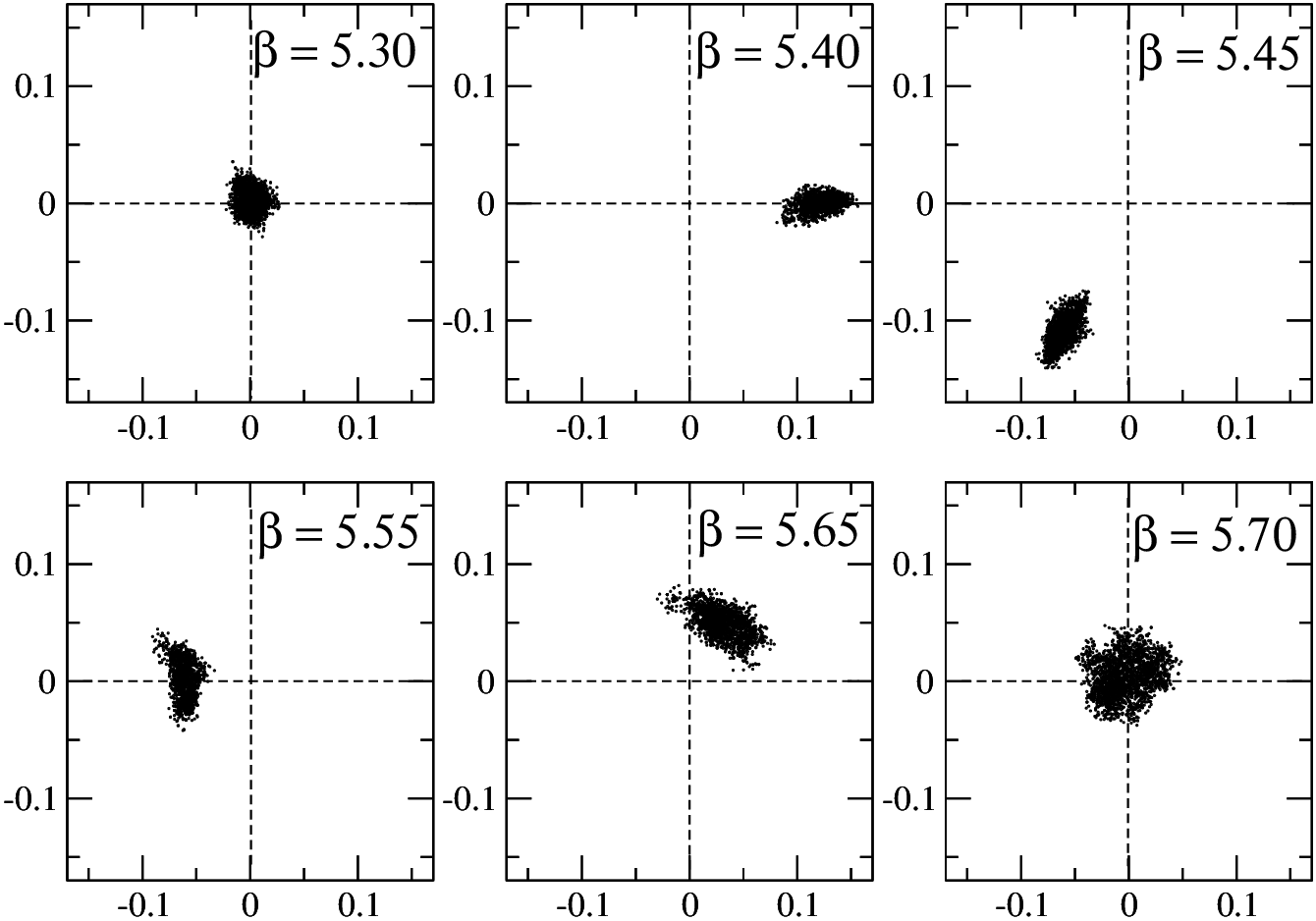}
\caption{
Same as in Fig.~\ref{distr1}, but for $a m = 0.02$.
}
\label{distr2} 
\end{figure}

In Figs.~\ref{distr1} and \ref{distr2} we show scatter plots of the 
distribution of $L_{(3)}$ in the complex plane, 
for $am = 0.1$ and $am = 0.02$ respectively, 
and for different values of $\beta$. In each case
four different phases are clearly distinguishable:

\begin{itemize}

\item
In a first phase at low $\beta$ values, correponding to large values
of $L_c$, $L_{(3)}$ is distributed around zero with a
vanishing expectation value. That corresponds to the usual confined,
center symmetric phase of QCD(Adj).\\

\item
As $\beta$ increases ($L_c$ is made shorter), the system passes 
into a phase with a broken center symmetry, with $L_{(3)}$ aligned 
along one of the center elements of SU(3). We call this phase
{\it deconfined}, since it is analogous to the usual 
high temperature phase of QCD.

\item
At very high values of $\beta$ (very small $L_c$) the system
is again in a center symmetric, confined-like phase, in agreement 
with the prediction of the one-loop potential in
Eq.~(\ref{potential}). 
Since this phase seems to be distinguished and separated 
from the low temperature confined
phase, we refer to it as {\it re-confined}.

\item
Finally, a new phase is present for values of $\beta$ ($L_c$) between
the deconfined and the re-confined phase, in which center symmetry 
is broken by a non-zero expectation value of $L_{(3)}$,
which is however oriented along directions rotated roughly by an angle
$\pi$, in the complex plane, with respect to those corresponding to 
center elements. This phase is completely analogous to that found
by the authors of Ref.~\cite{myog07} in their study of pure gauge
SU($N_c$) gauge theories with adjoint Polyakov line deformations,
and predicted again by the same authors, on the basis of the 1-loop
effective potential, also in the case of QCD(Adj) with p.b.c. and for some 
range of $m$ and $L_c$~\cite{myog09}.
We shall refer to this phase as a {\it split} or {\it skewed} phase, following 
the convention in Ref.~\cite{myog07}.

\end{itemize}

\begin{figure}[h!]
\includegraphics*[width=1.0\columnwidth]{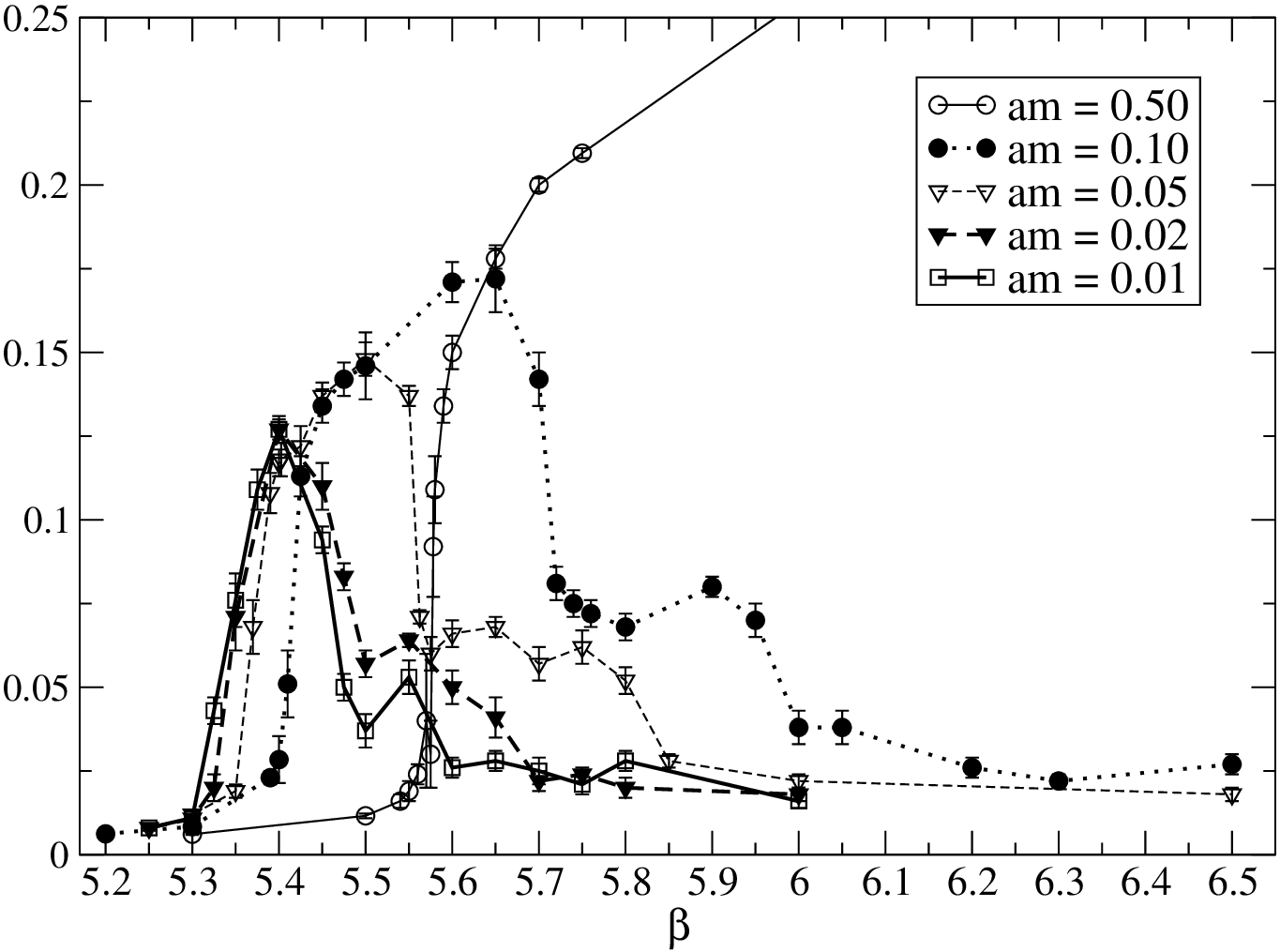}
\vspace{-0.cm}
\caption{
Average value of the Polyakov line modulus, on a $16^3\times 4$
lattice, as a function of $\beta$ and 
for different bare quark masses. 
}
\label{Polmod} 
\vspace{-0.cm}
\end{figure}

\begin{figure}[h!]
\includegraphics*[width=1.0\columnwidth]{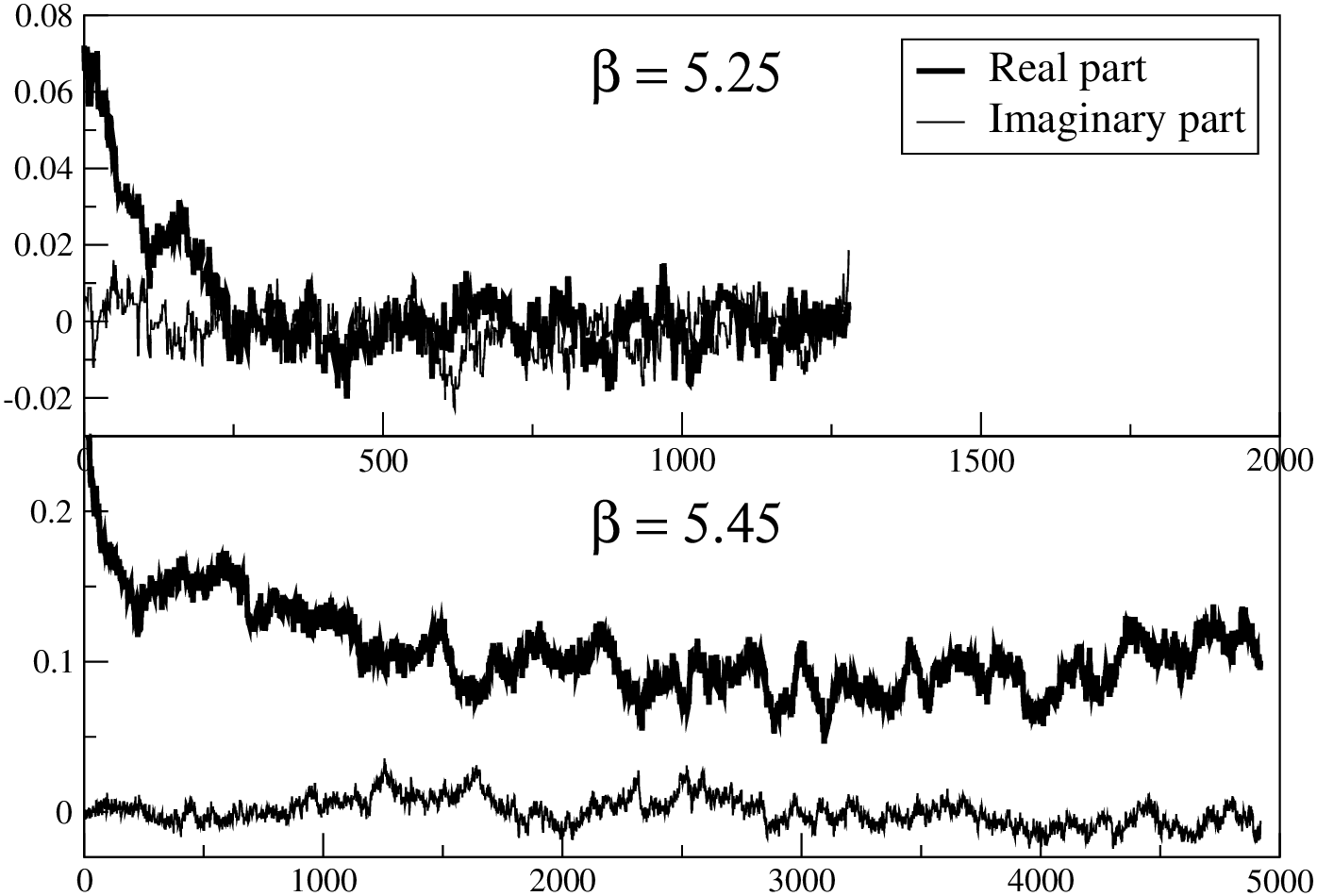}
\vspace{-0.cm}
\caption{
Time histories (in units of Molecular Dynamics trajectories and including part of the thermalization) of the real and imaginary part of $L_{(3)}$,
 for $am = 0.01$ and two different $\beta$ values in the 
confined and deconfined phase respectively.  
}
\label{history} 
\vspace{-0.cm}
\end{figure}

The transitions among the different four phases are also clear from 
Fig.~\ref{Polmod}, where we report the average value of 
$|L_{(3)}|$ as a function of $\beta$ for 
the different quark masses explored. In most cases
 the average modulus
is small at low $\beta$ (confined) phase; then starts growing at
a critical value $\beta_{c/d}$ separating the confined from the 
deconfined phase; a second critical value $\beta_{d/s}$ 
is reached, separating the deconfined from the split (skewed) phase,
at which the modulus drops roughly by a factor 3; finally
a third critical value $\beta_{s/r}$ is met where the system
goes into the new center symmetric phase with a small average
modulus for $L_{(3)}$. 
In the case of $a m = 0.5$ only two phases 
(confined and deconfined) are visible in Fig.~\ref{Polmod},
simply because of the limited range of $\beta$ shown in the figure.
The evidence for the existence of a split (skewed) phase is weaker
at the smallest quark mass explored, $a m = 0.01$.
Typical time histories for the 
real and imaginary parts of $L_{(3)}$ are shown in 
Fig.~\ref{history}, in particular for the confined and deconfined
phase at the lowest quark mass explored, $am = 0.01$.

It is interesting to notice that in each different phase the Polyakov
line modulus has a very small dependence on the quark mass, much
weaker than what observed in the case of finite temperature 
QCD(Adj) (see for instance results reported in
Ref.~\cite{Karsch,Engels}). In the present case the main effect of changing 
the quark mass seems that of moving the location of the phase transitions.

Locations of the critical couplings as a function of the bare quark
mass are reported in Table~\ref{betac}.
Regarding the order of the phase transitions, 
we have indications,
coming from abrupt jumps of observables or presence of
metastabilities, that they are all first order, at least for
$am \geq 0.05$. For the smallest quark masses, $am = 0.01$ 
and $am = 0.02$, transitions look smoother.
A deeper investigation and a finite size
scaling analysis would be needed to reach a definite conclusion:
that goes beyond our present purposes. However we notice  
that true phase transitions (continuous or discontinuous) 
should be found at points where the realization of the (exact) center
symmetry changes.

\begin{table}[t!]
\begin{center}
\begin{tabular}{|c||c|c|c|}
\hline $a m$ & $\beta_{\rm c/d}$ & $\beta_{\rm d/s}$ & $\beta_{s/c}$ \\
\hline
\hline 0.50  & 5.575(5) & 7.50(10)   & 8.50(20) \\
\hline 0.10  & 5.41(1)  & 5.71(1)  & 6.00(5)\\
\hline 0.05  & 5.37(1)  & 5.56(1) & 5.82(3)\\
\hline 0.02  & 5.33(1)  & 5.485(10)   & 5.65(3)\\
\hline 0.01  & 5.32(1)  & 5.465(10)   &  5.55(5)  \\
\hline
\end{tabular}
\end{center}
\caption{Critical values of $\beta$ as a function of the bare quark
  mass for the transition from the confined to the deconfined phase
($\beta_{\rm c/d}$), from the deconfined to the split (skewed) phase
($\beta_{\rm d/s}$), and from the 
split to the weak coupling confined (re-confined) phase
  ($\beta_{s/c}$). 
In the case of $am = 0.01$ the split phase is absent 
and $\beta_{\rm d/s}$ actually indicates the
transition from the deconfined to the re-confined phase.
\label{betac}}
\end{table}

\begin{figure}[t!]
\includegraphics*[width=1.0\columnwidth]{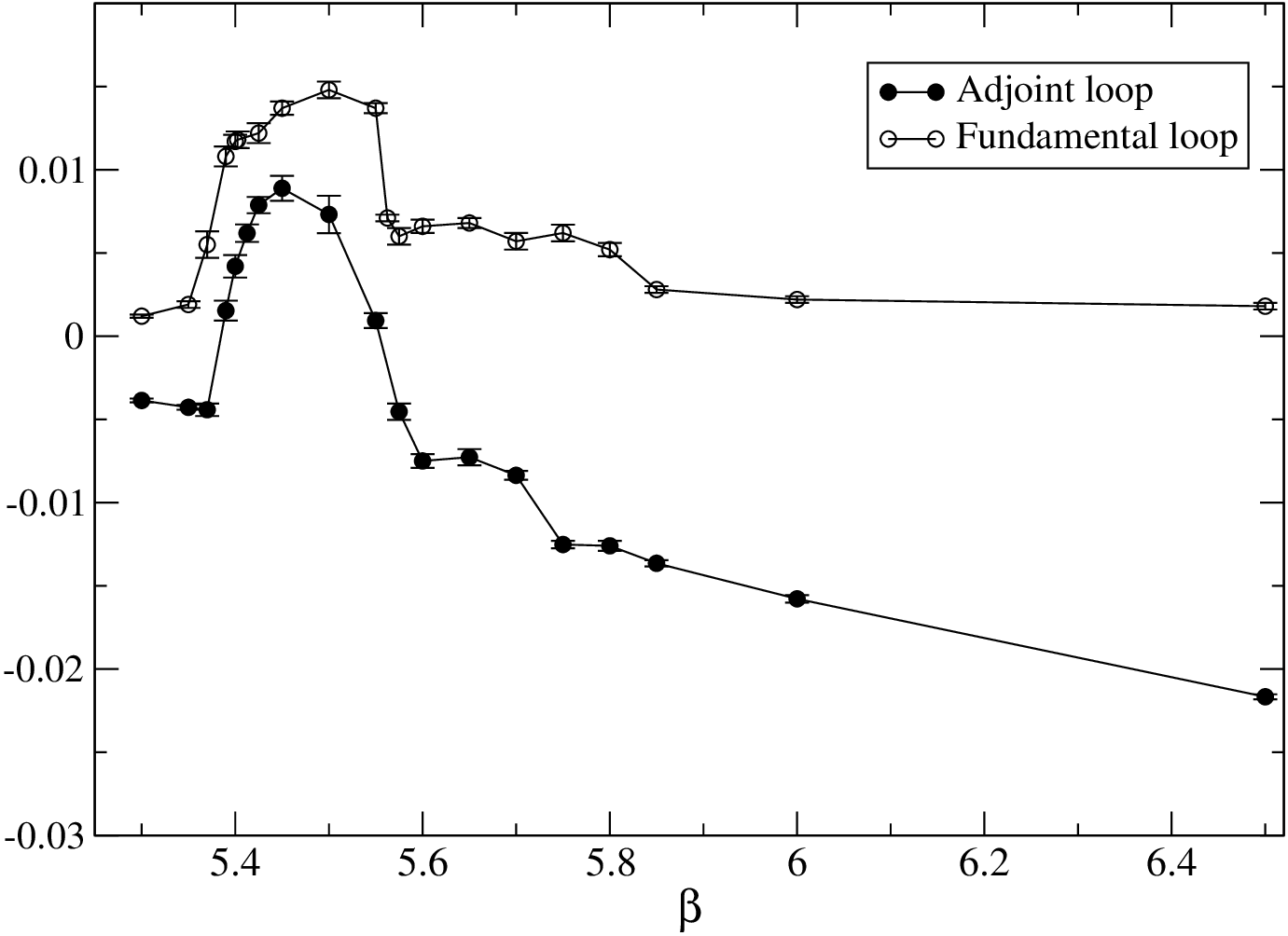}
\vspace{-0.cm}
\caption{
Adjoint Polyakov loop compared to the modulus of the fundamental loop
(divided by a factor ten to fit in the figure), as a function of
$\beta$ for $a m = 0.05$ on a $16^3 \times 4$ lattice.
}
\label{adjointloop} 
\vspace{-0.cm}
\end{figure}

Let us now discuss our results.
The confining behaviour observed at large $L_c$ (low $\beta$)
is in agreement
with the usual infinite volume behaviour of QCD(Adj). The 
confining behaviour observed at small $L_c$ (high
$\beta$) is in agreement with the prediction of the 
one-loop effective potential in Eq.~(\ref{potential}):
fermions contribute with a repulsive, disordering interaction
for the Polyakov line eigenvalues, which is opposite and wins
over the ordering gluon contribution. What is non-trivial 
is the behaviour observed at intermediate values
of $L_c$, where it is clear that higher order and finite quark 
mass corrections to the one-loop effective potential
play a significant role. In the deconfined phase those corrections
result in an overall attractive interaction for the 
eigenvalues which orders the Polyakov line along a center
element. In the case of the split (skewed) phase instead, the free energy
minimum is apparently reached as two eigenvalues coincide and 
the third one gets a phase $\pi$ with respect to them,
as in the following SU(3) matrix
\beq
e^{i k / N_c}
\left( \begin{array}{rrr}
-1 & 0 & 0 \\
0 & -1 & 0 \\
0 & 0 & 1 \end{array} \right)
\label{skew}
\eeq
That is not a minimum for the potential in Eq.~(\ref{potential}),
which is instead minimized (in the case of p.b.c. and for $N_f > 1$) by the traceless matrix 
${\rm diag}(1,e^{2 i \pi/3},e^{-2 i \pi/3})$, however it has been
verified in Ref.~\cite{myog07} that mass corrections to the 
1-loop potential
may move the minimum to group elements like that in Eq.~(\ref{skew}).
The trace of such group elements is 1/3 with respect to that of center
elements: that roughly explains the drop observed for 
the modulus of $L_{(3)}$ at $\beta_{d/s}$.

It is also interesting to look at the behaviour of the adjoint
Polyakov loop, which is shown in Fig.~\ref{adjointloop} for
the case $am = 0.05$: it is positive only in the deconfined phase
and negative outside. Since ${\rm Tr}_{(8)} = |{\rm Tr}_{(3)}|^2 - 1$,
the large negative values obtained for the average adjoint Polyakov loop
in the re-confined phase can be interpreted as the fundamental loop
becoming traceless point by point in the weak-coupling limit, 
in agreement with the 1-loop potential in Eq.~(\ref{potential}),
while in the strong coupling regime the fundamental trace averages
to zero mainly because of long range disorder.

\begin{figure}[h!]
\includegraphics*[width=1.0\columnwidth]{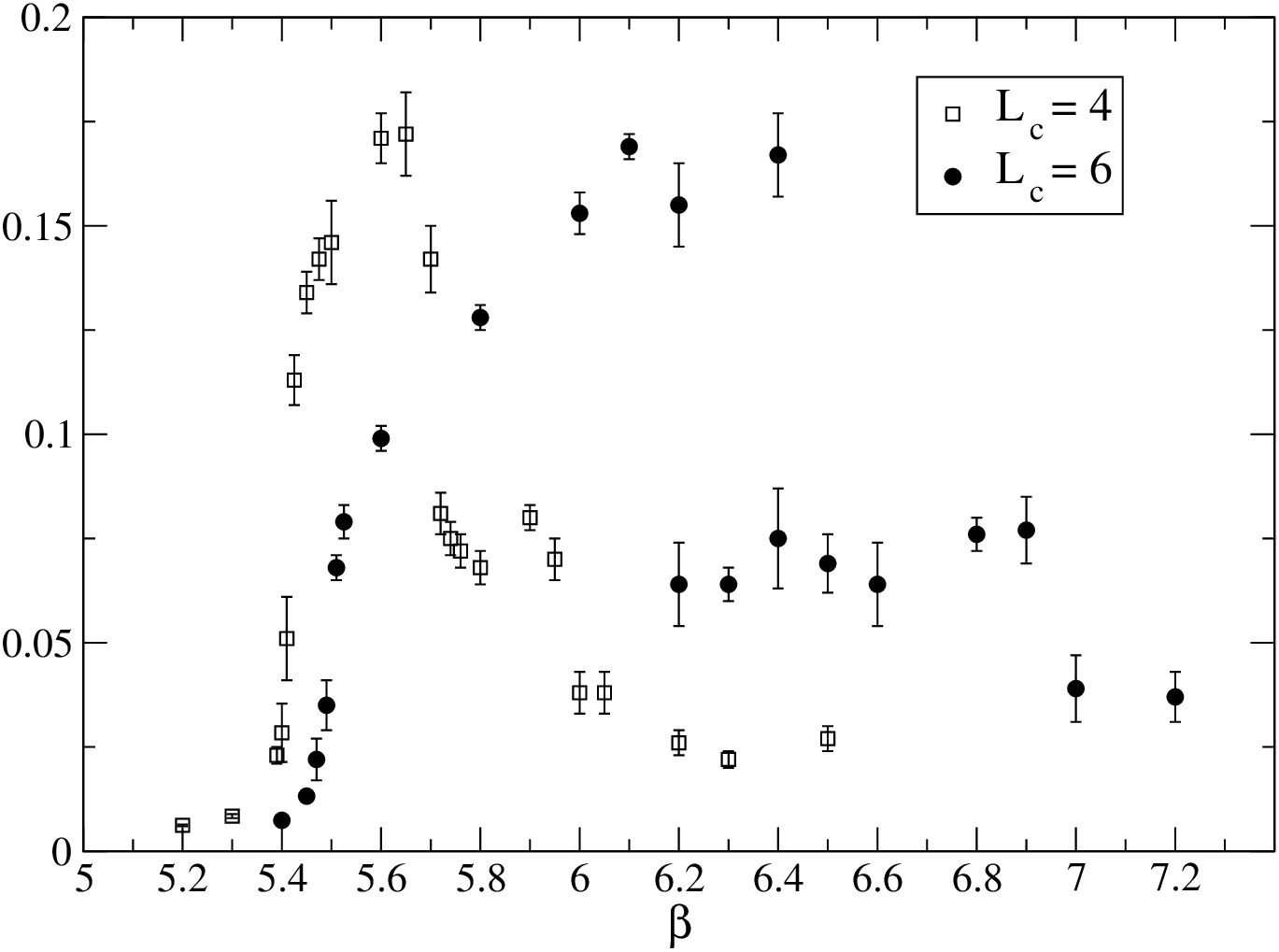}
\vspace{-0.cm}
\caption{
Comparison of results obtained for the 
average value of the Polyakov line modulus as a function
of $\beta$ at $am = 0.10$ and on two different lattices,
$16^3\times 4$ and $16^3\times 6$.
}
\label{nt6nt4} 
\vspace{-0.cm}
\end{figure}

In the case of bare quark mass $a m = 0.10$ we have repeated 
our runs on a $16^3 \times 6$ lattice, i.e. taking
$\hat L_c = 6$ sites in the compactified direction. Results for the modulus of 
$L_{(3)}$ are reported in Fig.~\ref{nt6nt4}, where they are compared
with results obtained on the $16^3 \times 4$ lattice. The phase
structure does not change and all the critical couplings move 
to larger values: that is consistent with the possible existence of 
a continuum limit for the critical values of the 
compactified dimension $L_c = \hat L_c a(\beta,am)$. 
In particular the deconfinement critical coupling 
moves from $\beta_{c/d} (\hat L_c = 4) = 5.41(1)$ to
$\beta_{c/d}(\hat L_c = 6) = 5.50(1)$: that is consistent
with a constant critical length $L_{c/d} = \hat L_c a(\beta_{c/d})$,
since using the 2-loop $\beta$-function
\beq
a\Lambda_L \approx R(\beta) = \left({ \beta \over 6b_0} \right)^{b_1/2b_0^2}
\exp \left(- {\beta \over 12 b_0} \right) 
\label{beta}
\eeq
with $b_0$ and $b_1$ given by~\cite{Caswell}
\beq
b_0 = {3 \over 16 \pi^2}~, \quad b_1 = -  {90 \over (16\pi^2)^2}~,
\label{coeff}
\eeq
we get 
$${a(\beta_{c/d} (\hat L_c = 4))\over a(\beta_{c/d} (\hat L_c = 6)}
\simeq 1.6(2)\ .$$
$\beta_{d/s}$ and $\beta_{s/r}$ show a larger change when going 
from $\hat L_c = 4$ to $\hat L_c = 6$, which cannot be interpreted in
terms of the 2-loop $\beta$-function. This is due to the fact that
we are not working at fixed physical quark mass, but instead at 
fixed $a m$, meaning larger and larger $m$ as $\beta$ increases:
as we increase $\hat L_c$ we have to increase $\beta$ to make 
$a$ smaller and keep $L_c = a \hat L_c $ fixed, but at the same 
time, if $am$ is fixed, 
also $m = (am)/a$ increases meaning that also the physical value of 
$L_c$ changes. While the critical $L_{c/d}$ corresponding 
to deconfinement changes smoothly and stays 
finite as the quenched limit $m \to \infty$ is approached, 
so that this effect may be small,
the other two transitions must disappear in the same limit,
meaning a steepest dependence of the physical critical
dimension on $m$ (hence on $\beta$).
We notice that for $\hat L_c = 6$ the transition from the
deconfined to the split (skewed) phase is accompanied by strong
metastabilities, testified by different simulations at equal
values of $\beta$ staying in different phases for thousands of
trajectories (see Fig.~\ref{nt6nt4} where the results of those
simulations are reported for $\beta = 6.2$ and $\beta = 6.4$).

\begin{figure}[t!]
\includegraphics*[width=1.0\columnwidth]{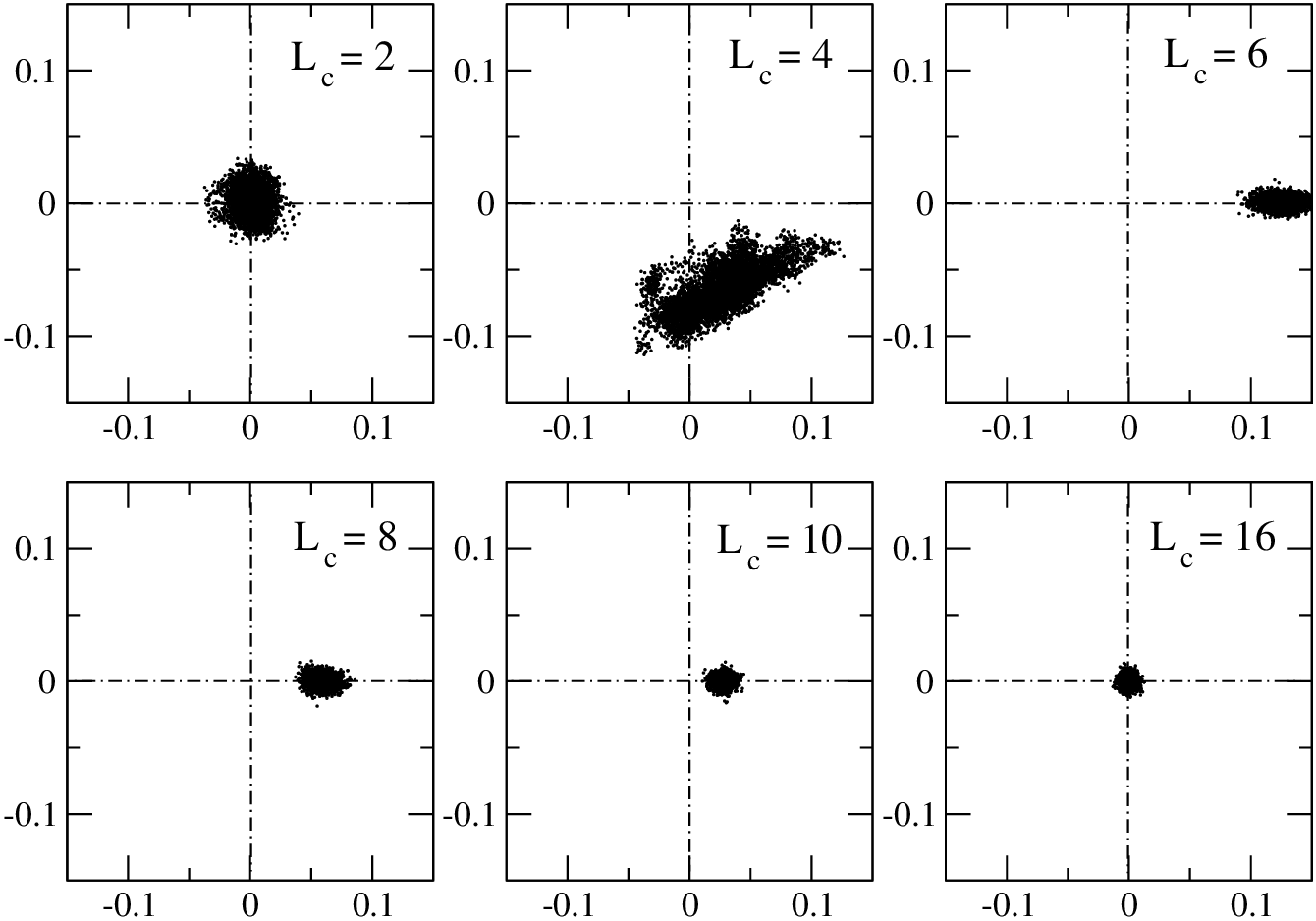}
\vspace{-0.cm}
\caption{
Scatter plots of the distribution of $L_{(3)}$
in the complex plane on a $16^3\times \hat L_c$ lattice at fixed quark mass
and UV cutoff ($\beta = 5.75$ at $am = 0.10$) for various 
values of the compactified dimension $\hat L_c$.
}
\label{lcvardistr} 
\vspace{-0.cm}
\end{figure}

In order to get an idea of the scale separation between the different
phases we have also performed simulations at fixed quark mass
and ultraviolet (UV) cutoff, and variable $\hat L_c$ (only even values
of $\hat L_c$ can be explored because of the staggered fermion
discretization). In particular in Fig.~\ref{lcvardistr} 
we show 
the distribution of $L_{(3)}$ in the complex plane and 
its average modulus for different value of $\hat L_c$ on a
$16^3 \times \hat L_c$ lattice, at fixed 
bare quark mass $am = 0.1$ and inverse gauge coupling $\beta = 5.75$.
The system is in the confined, center symmetric phase for 
$\hat L = 16$, where $\langle L_{(3)} \rangle =
(0.0001(2),0.0002(2))$; it is still marginally confined for 
$\hat L_c = 14$, where 
$\langle L_{(3)} \rangle = (0.0005(2),0.0001(2))$, while as
the compactified dimension is squeezed further a finite Polyakov line
expectation value develops already for $\hat L_c = 12$, where 
$\langle L_{(3)} \rangle = (0.0066(8),-0.0008(8))$: this is clearly
different from zero, even if quite small 
due to strong UV noise. The system stays in the deconfined
phase till $\hat L_c = 6$, while it is clearly in the split phase for 
$\hat L_c = 4$. Finally, for the shortest
compactified dimension explored, $\hat L_c = 2$, it is 
in the center symmetric, re-confined phase. We conclude
that, in this case, the two center
symmetric, confined phases, taking place at large or small values of the 
compactified dimension, are not connected: they have boundaries which 
are separated by a scale factor $\sim$ 3-4.

\subsection{Approaching the chiral limit}

Our results do not support the conjecture about volume independence of 
QCD(Adj), since for all explored masses the two center symmetric phases 
taking place for large and small values of $L_c$ do not correspond to a unique 
confined phase, but are instead separated by different phases in which center
symmetry is spontaneously broken. However a very
important issue is whether this is true for arbitrarily small masses.

While our computational resources do not allow us to perform extensive
simulations for smaller quark masses, we can try to extrapolate
to the chiral limit our results for the critical values 
of $\beta$ obtained on the $16^3 \times 4$ lattice and 
reported in Table~\ref{betac}. We have therefore fitted
the critical values of $\beta$ obtained for each different
phase transition according to 
\beq
\beta_{crit} = \beta_{crit}^\chi + c_1 a m + c_2 (a m)^2
\label{quadratic}
\eeq
where $\beta_{crit}^\chi$ indicates the critical coupling in the
chiral limit. 

\begin{figure}[t!]
\includegraphics*[width=1.0\columnwidth]{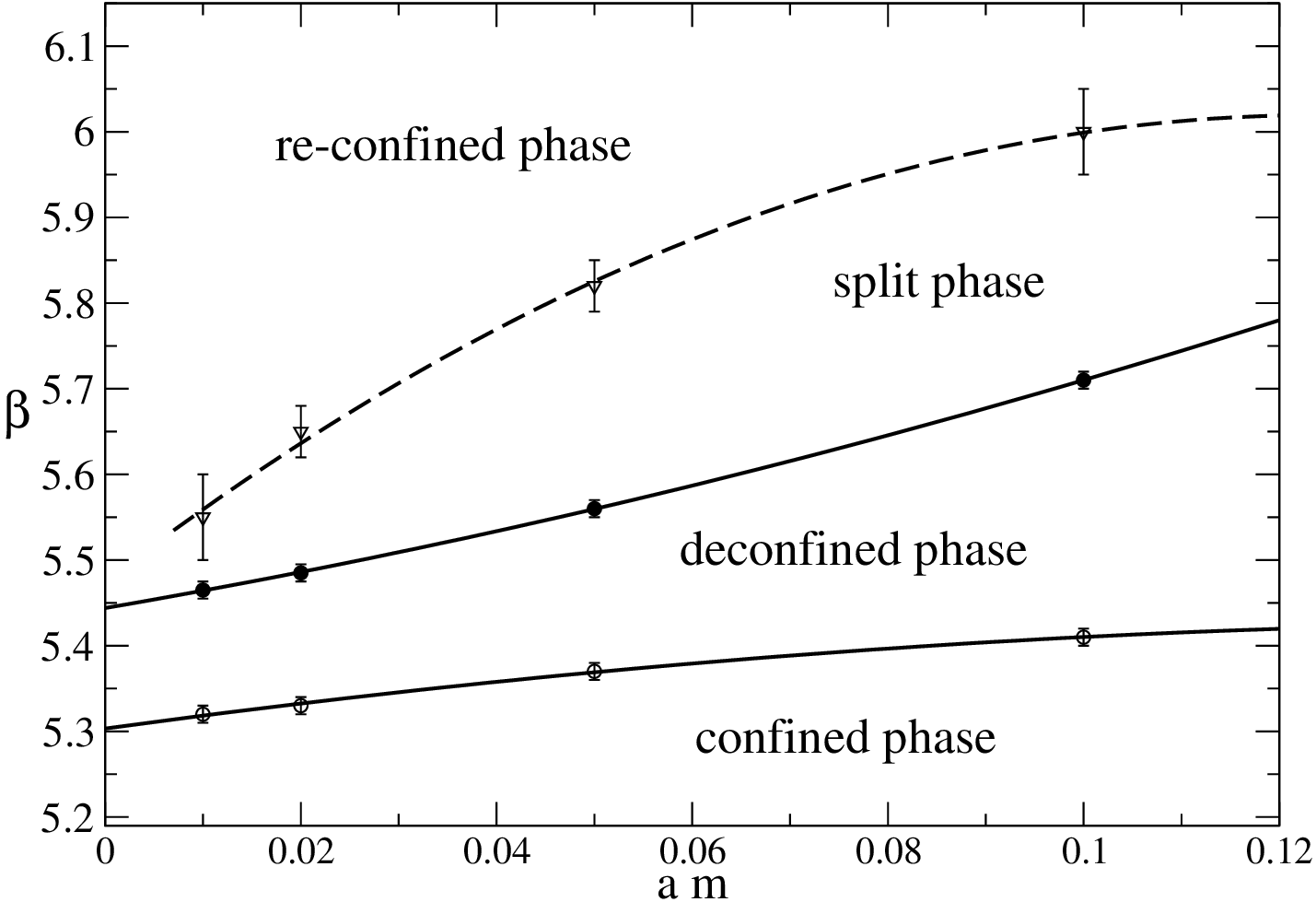}
\vspace{-0.cm}
\caption{
Possible sketch of the phase diagram in the $\beta$-$am$ plane and approaching
the chiral limit, obtained by means of quadratic interpolations of the 
critical couplings as in Eq.~(\ref{quadratic}).
}
\label{diagram} 
\vspace{-0.cm}
\end{figure}

Results are reported in Fig.~\ref{diagram}. The split phase 
could disappear in the chiral limit, and already at $am = 0.01$ 
the evidence for its existence is weaker.
The values of the critical
couplings extrapolated to the chiral limit are the following:
$\beta_{c/d}^\chi = 5.310(15)$ 
for the transition from the confined to the deconfined phase, 
and 
$\beta_{d/s}^\chi = 5.444(15)$
for the transition from the deconfined to the split 
(or re-confined) phase.

These results provide partial evidence for
a deconfined phase surviving in the chiral limit. 
The two confined phase are not connected but still
separated, even in the chiral limit, by a scale factor which is 
estimated, on the basis of the 2-loop  $\beta$-function given in
Eq.~(\ref{beta}), to be roughly 2. We would like to stress that this
is only the result of an extrapolation to the chiral limit of the
critical couplings $\beta_{c/d}$ and $\beta_{d/s}$ delimiting the 
deconfined phase in the phase diagram in Fig.~\ref{diagram}: there is still no definite evidence
that the deconfined phase actually extrapolates to $am = 0$.

\begin{figure}[t!]
\includegraphics*[width=1.0\columnwidth]{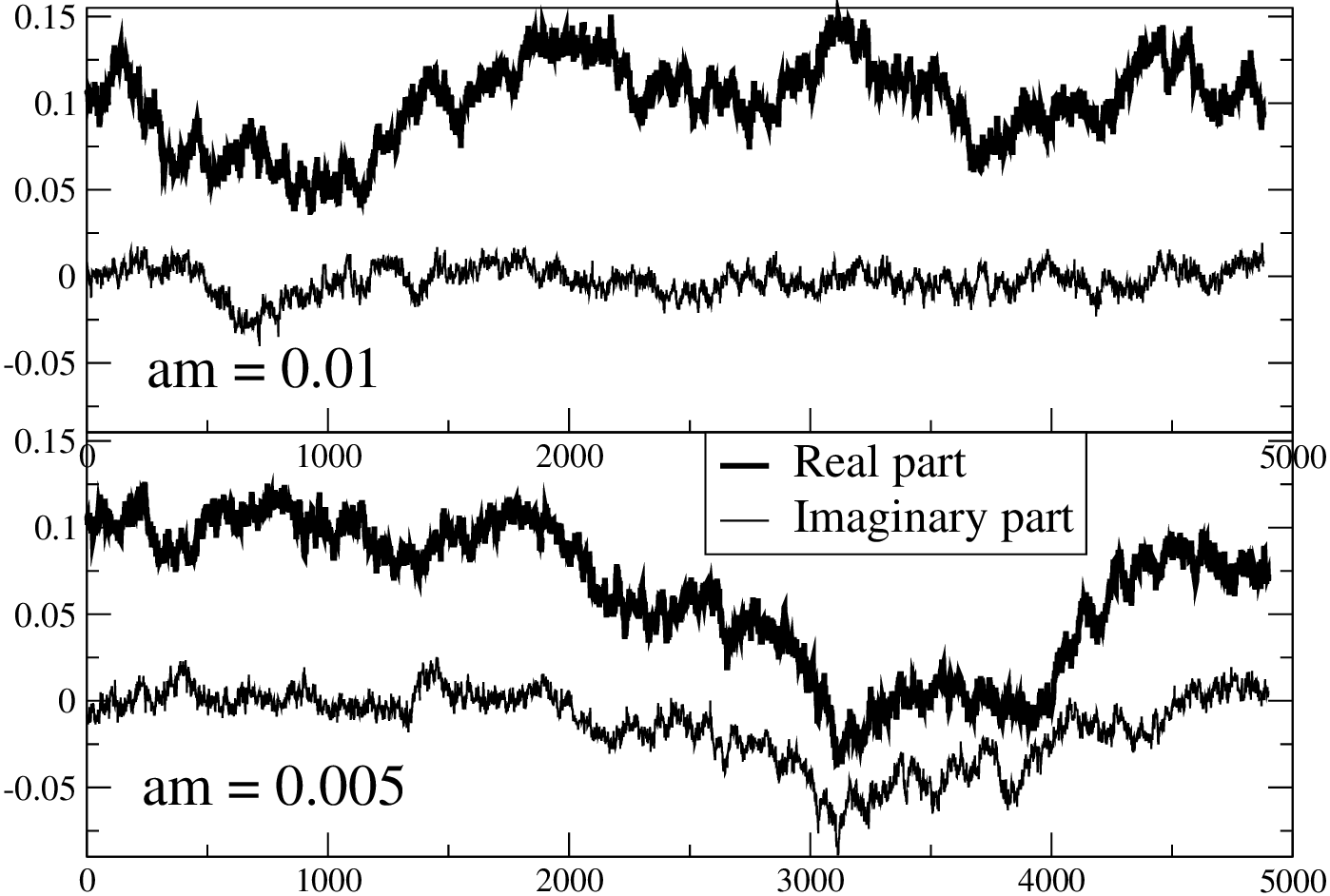}
\vspace{-0.cm}
\caption{
Time histories, in units of Molecular Dynamics trajectories, of the real and imaginary part of $L_{(3)}$,
 for $\beta = 5.4$ and two different values of $a m$. 
}
\label{smallmass} 
\vspace{-0.cm}
\end{figure}

In order to get more information on this issue we have performed
a numerical simulation at a smaller value of the quark mass, $am = 0.005$, and
for a single value of $\beta$ chosen well within the deconfined region
in Fig.~\ref{diagram}, $\beta = 5.40$. In Fig.~\ref{smallmass} we report
the Monte-Carlo time histories obtained for the real and imaginary part of the Polyakov
loop respectively at $a m = 0.01$ and $a m = 0.005$. It is apparent from the figure that
while for $am = 0.01$ the system stays quite stably in the deconfined phase, for the lower
mass it spends part of the simulation time into a phase with a lower value of the Polyakov
loop (seemingly a split phase). More extensive (not affordable) simulations would be necessary
to clarify the issue, but we can at least say that some metastable behaviour is present
at this mass which may suggest an even less trivial phase structure close to the chiral limit.

\subsection{Chiral properties}

It is interesting to study the fate of chiral symmetry and
its connection with the different observed realizations of center
symmetry in this theory. 
The study of finite temperature QCD(Adj) has shown that, unlike
the case of ordinary QCD, phases may exist 
in which deconfinement is not accompanied by chiral symmetry 
restoration~\cite{Karsch,Engels,Cossu}. In the present case, since
a chirally symmetric phase is expected anyway in the weak 
coupling regime (i.e. for short enough $L_c$), a new exotic phase could
exist in which both chiral symmetry and center symmetry are not
spontaneously broken~\cite{Unsal:2007vu,Unsal:2007jx}.

\begin{figure}[t!]
\includegraphics*[width=1.0\columnwidth]{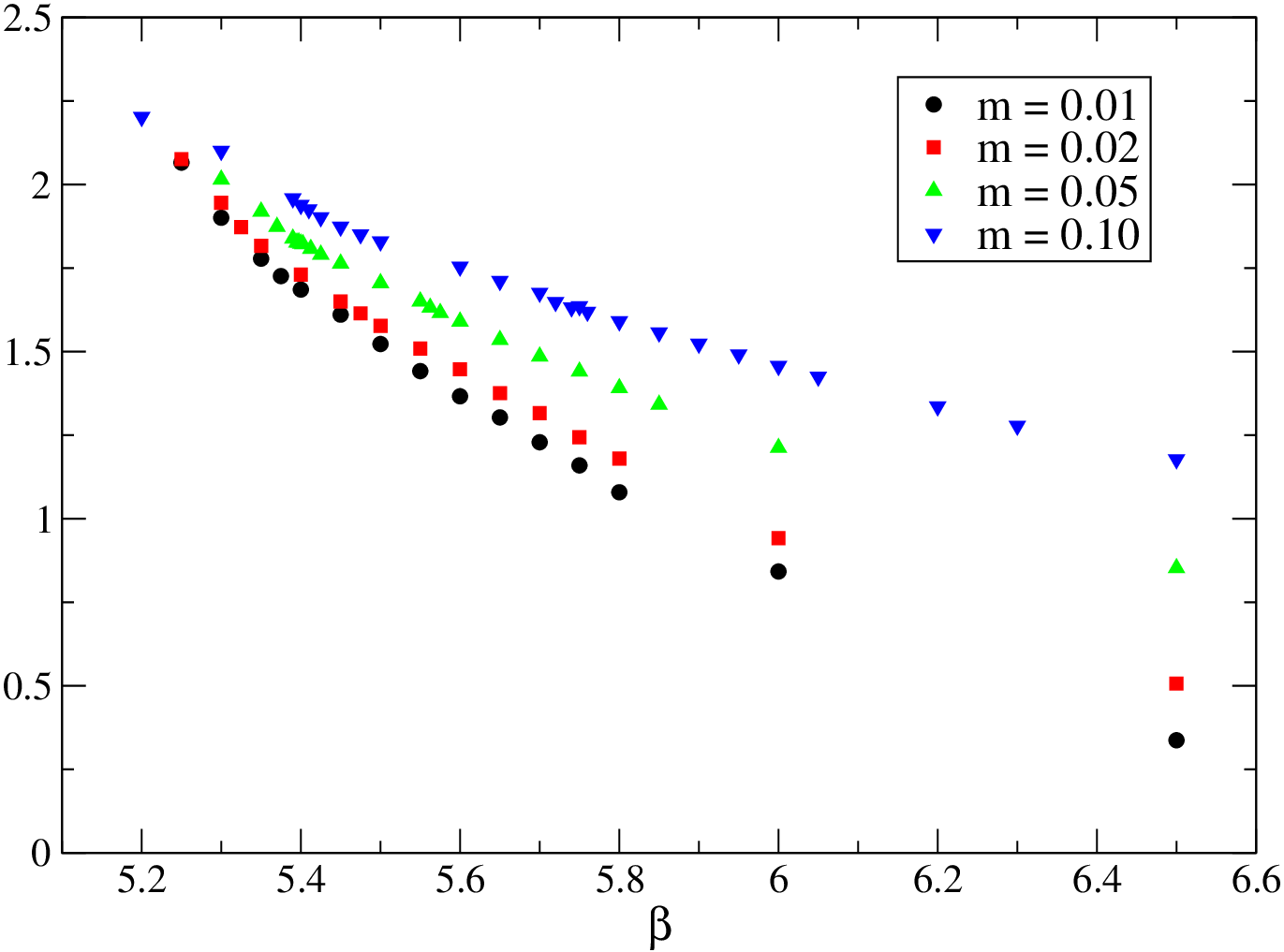}
\vspace{-0.cm}
\caption{
Chiral condensate as a function of $\beta$ for different values of the
bare quark mass on the $16^3 \times 4$ lattice.
}
\label{chiral_beta} 
\vspace{-0.cm}
\end{figure}

\begin{figure}[t!]
\vspace{0.5cm}
\includegraphics*[width=1.0\columnwidth]{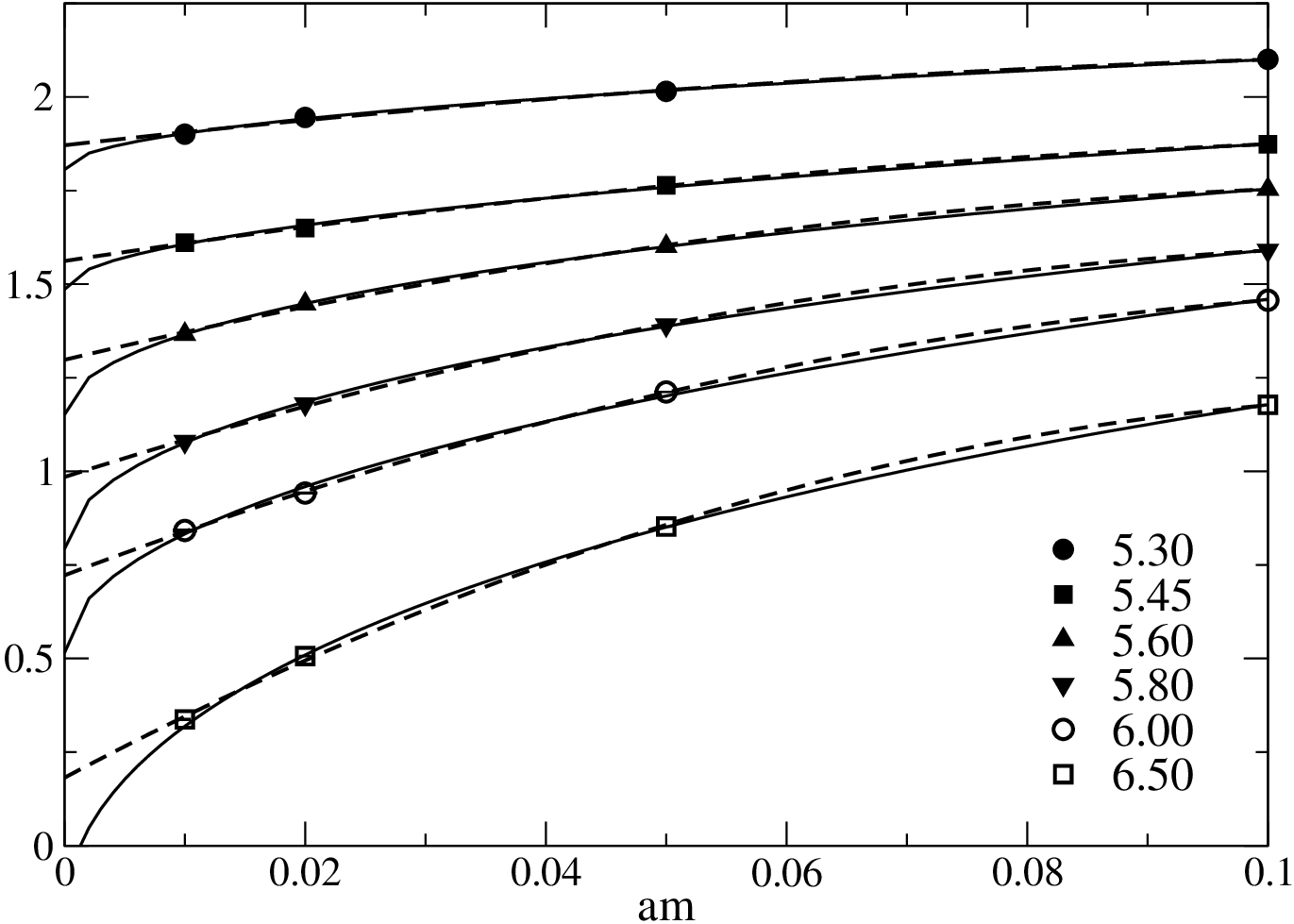}
\caption{
Chiral condensate as a function of the quark mass and quadratic extrapolations
to the chiral limit including (continuous) or not including (dashed) a 
term proportional to $\sqrt{am}$, 
for different values of the gauge coupling on the
$16^3 \times 4$ lattice.
}
\label{chiral_extra} 
\vspace{-0.cm}
\end{figure}

In Fig.~\ref{chiral_beta} we show the behaviour of the chiral
condensate as a function of $\beta$ for different quark masses, as
obtained on the $16^3\times 4$ lattice, while
in Fig.~\ref{chiral_extra} we show the extrapolations of the 
same chiral condensate values 
to the chiral limit for different values of $\beta$
which cover all the possible different phases described by the
Polyakov line. Also in this case we have used quadratic 
extrapolations 
\beq
\langle \bar\psi \psi \rangle (am) = 
\langle \bar\psi \psi \rangle (0) + a_1 a m + a_2 (a m)^2
\label{condextra}
\eeq
which are shown as dashed lines in Fig.~\ref{chiral_extra}, together
with similar extrapolations, including also a term proportional to 
$\sqrt{am}$ (continuous lines), analogous to those used in Ref.~\cite{Karsch}.

\begin{figure}[t!]
\includegraphics*[width=1.0\columnwidth]{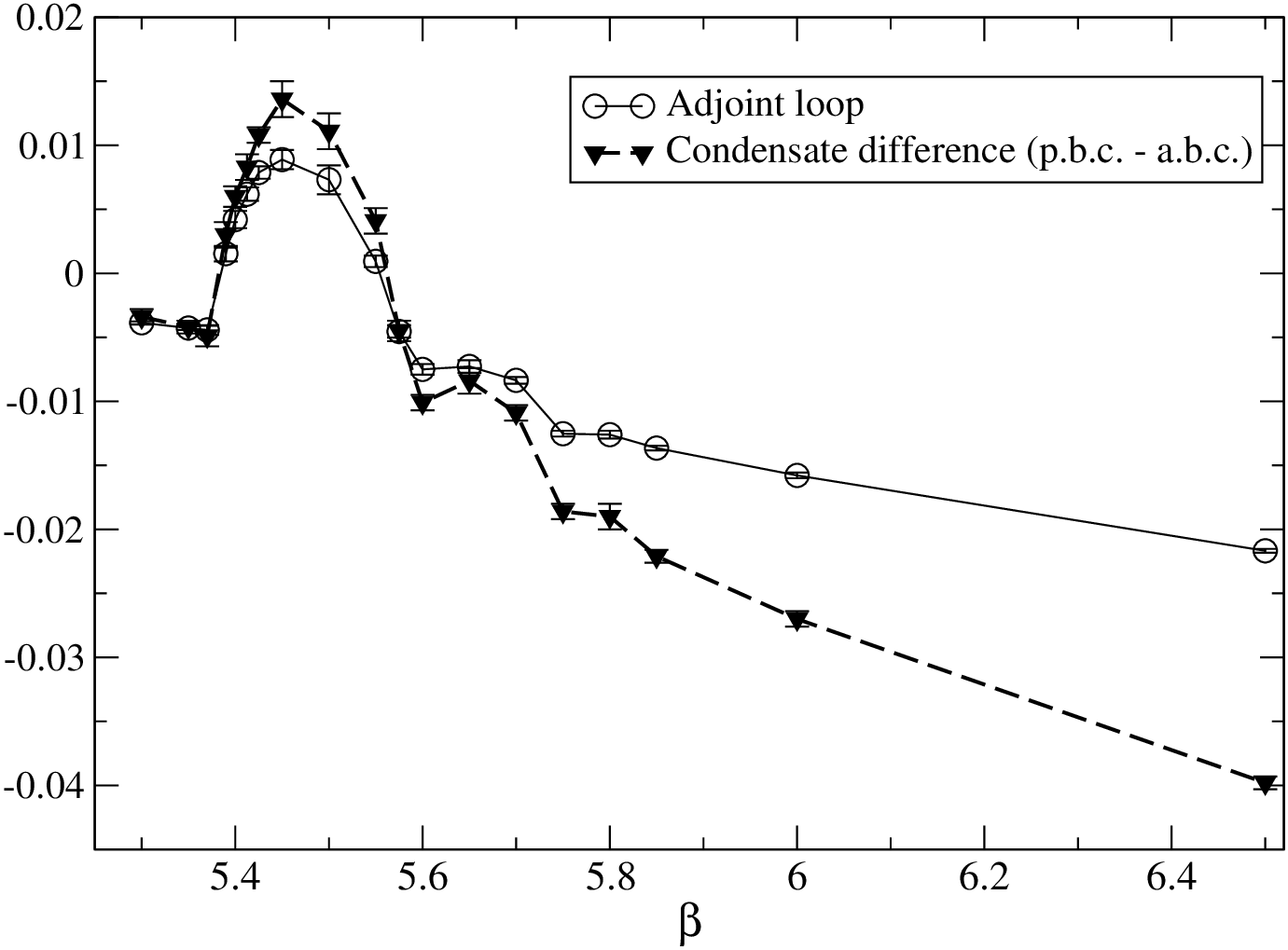}
\vspace{-0.cm}
\caption{
Comparison between $(\langle \bar\psi \psi\rangle_m^{(0)} -  \langle \bar\psi \psi\rangle_m^{(\pi)})$
(see text) and the adjoint Polyakov loop for $am = 0.05$ on a $16^3\times 4$ lattice.
}
\label{chiralprop} 
\vspace{-0.cm}
\end{figure}

Fig.~\ref{chiral_extra} shows that, independently of the extrapolation
used, the chiral condensate
always extrapolates to a non-zero value, as $am \to 0$, in all the
range of $\beta$ values going at least up to $\beta = 6.0$. That means that 
the different transitions, corresponding to different realization of
center symmetry, do not affect chiral symmetry, which remains
spontaneously broken.

Instead at the highest value explored, $\beta = 6.50$, 
there is some evidence that the
chiral condensate may extrapolate to zero and chiral symmetry be restored:
that would fit the expectation of a chirally symmetric weak coupling
phase. However we cannot draw a definite conclusion from present
results: at this value of $\beta$ the lattice spacing is very small
and even $a m = 0.01$ corresponds to a large quark mass,
so that an extrapolation to zero quark mass may be debatable;
moreover one should also properly take into account the running of 
the adimensional quark condensate measured on the lattice 
as the continuum limit $a \to 0$ is approached.

Therefore, the only reasonable conclusion that we can draw from
our results is that the expected weak coupling restoration of 
chiral symmetry may happen at values of the compactified dimension
much shorter (at least one order of magnitude, as roughly estimated
on the basis of the 2-loop $\beta$-function) than those at which
center symmetry transitions take place.

It is interesting to ask if the information about the change in the realization
of center symmetry is transcripted in some other way into the chiral properties of the 
theory. A possible candidate to give an answer is the dual chiral condensate 
introduced in Ref.~\cite{dualcond} and defined as
\beq
\Sigma_1 = -\frac{1}{2\pi} \int_0^{2\pi} d\varphi\ e^{-i\varphi} \langle \bar\psi \psi\rangle_m^{(\varphi)}
\label{dualcondensate}
\eeq
where $\langle \bar\psi \psi\rangle_m^{(\varphi)}$ is the chiral condensate measured with
an assigned phase $\exp(i \varphi)$ for the fermionic boundary conditions in the compactified dimension
($\varphi = 0$ for p.b.c. and $\varphi = \pi$ for a.b.c.), and $\Sigma_1$ is nothing but its
Fourier transform with respect to the phase.  
It can be easily shown, e.g. by a loop expansion of the fermionic determinant, that 
in the infinite quark mass limit the dual condensate becomes the usual Polyakov loop 
(adjoint Polyakov loop in this case), hence it is a natural chiral quantity which could
be sensible also to the realization of center symmetry, as the adjoint loop is 
(see Fig.~\ref{adjointloop}). As a very rough estimate of $\Sigma_1$, we have considered 
the quantity $(\langle \bar\psi \psi\rangle_m^{(0)} -  \langle \bar\psi \psi\rangle_m^{(\pi)})$,
i.e. the difference in the condensate measured using p.b.c. and a.b.c. respectively (that means
only a change in the observable, which is measured on the same configurations produced with
dynamical fermions having p.b.c.). In Fig.~\ref{chiralprop} we report results obtained
for $am = 0.05$: as can be appreciated this difference follows quite closely the behaviour 
of the adjoint Polyakov loop, changing its sign in the deconfined phase.
This is a confirmation, in a theory with a highly non-trivial phase structure,
that the realization of center symmetry is reflected in the way 
the chiral condensate depends on the fermion boundary conditions~\cite{Bilgici}.

\section{Conclusions}

We have studied QCD with three colors,
2 flavors of Dirac (staggered) fermions in the adjoint representation
and one of the spatial dimensions compactified with periodic boundary
conditions for fermions.
We have shown that 4 different phases are explored as 
the size $L_c$ of the compactified dimension is varied, corresponding
to different realizations of center symmetry. In particular 
two center symmetric,
confined phases exist, for small and large values of $L_c$, separated
by two phases (deconfined and split phase) 
in which center symmetry is spontaneously broken
in two different ways. We have some partial evidence
that the deconfined phase could persist and
separate the two confined phases also in the chiral limit. 
This conclusion is based on an extrapolation to the chiral limit
of the critical lines determined for quark masses down to $a m = 0.01$.
On the other hand a direct simulation at a lower mass, $am = 0.005$, 
and at $\beta = 5.4$, which is inside the deconfined phase according
to the extrapolations above, has shown some sign of metastability which 
could signal a less trivial phase structure close to the chiral limit.
More extensive simulations at lower masses, which are not affordable by our present
computational resources, would be needed to better clarify the issue.

We have investigated the chiral properties of the theory and have provided evidence that 
chiral symmetry is not restored until very small values
of $L_c$, at least one order of magnitude shorter than those at which
center symmetry transitions take place. On the other hand, we have
verified that the realization of center symmetry is reflected 
in the way the chiral condensate depends on the fermion boundary conditions,
thus confirming similar results obtained in Ref.~\cite{Bilgici}.

Our results may be relevant for a recently proposed conjecture
about volume independence of QCD(Adj) in the large $N_c$
limit~\cite{KUY}: in particular the presence of a deconfined phase separating 
the large and small volume confined phases would not support
such conjecture. 
Of course, apart from the fact that the extrapolation to the chiral
limit of our findings should be further investigated,
our results could be relevant just to the particular case
studied here and to the discretization setup used.
It is therefore essential to perform further studies. 
In particular one should verify our results as the continuum
limit is approached, also making use of improved actions or 
different discretization setups, based for instance on Wilson 
instead of staggered fermions. Finally, one should consider 
that of course the situation could be different for larger values of $N_c$.

\section*{Acknowledgments}
We thank A.~Patella, C.~Pica and M.~Unsal for 
very useful discussions. This work has been started during the 
workshop ``Non-Perturbative Methods in Strongly Coupled Gauge
Theories'' at the Galileo Galilei Institute (GGI) in Florence.

\end{document}